\documentclass[conference]{IEEEtran}
\usepackage{tikz}
\usepackage{import}
\usepackage{ifthen}
\usepackage{amsmath}
\usepackage{tabularx}
\usepackage{array}
\usepackage{pgfplots}
\usepackage{float}
\usepackage[binary-units=true]{siunitx}
\usepackage{algorithm}
\usepackage{algpseudocode}
\usepackage[position=b]{subcaption}
\usepackage{lipsum}
\usepackage{pifont}
\usepackage{xspace}
\usepackage{times}

\usepgfplotslibrary{statistics}
\usepgfplotslibrary{groupplots}
\usetikzlibrary{shapes,arrows}

\newcommand{\bcOne}{\ding{182}}
\newcommand{\bcTwo}{\ding{183}}
\newcommand{\Hydra}{Scylla\xspace}

\pdfoutput=1

\begin{document}
\title{Execution Integrity with In-Place Encryption}

\author{
    \IEEEauthorblockN{
        Dean Sullivan\IEEEauthorrefmark{1},
        Orlando Arias\IEEEauthorrefmark{1},
        David Gens\IEEEauthorrefmark{2},
        Lucas Davi\IEEEauthorrefmark{3},
        Ahmad-Reza Sadeghi\IEEEauthorrefmark{2},
        Yier Jin\IEEEauthorrefmark{1}}
    \IEEEauthorblockA{\IEEEauthorrefmark{1}University of Central Florida, USA
    \\\{dean.sullivan, oarias, yier.jin\}@eecs.ucf.edu}
    \IEEEauthorblockA{\IEEEauthorrefmark{2}Technische Universit{\"a}t
    Darmstadt, Germany
    \\\{david.gens, ahmad.sadeghi\}@trust.tu-darmstadt.de}
    \IEEEauthorblockA{\IEEEauthorrefmark{3}University of Duisburg-Essen,
    Germany
    \\\{lucas.davi\}@wiwinf.uni-due.de}
}

\maketitle

\begin{abstract}
	Instruction set randomization (ISR) was initially proposed with the main goal
of countering code-injection attacks. However, ISR seems to have lost its
appeal since code-injection attacks became less attractive because protection
mechanisms such as data execution prevention (DEP) as well as code-reuse attacks
became more prevalent. 

In this paper, we show that ISR can be extended to also protect against
code-reuse attacks while at the same time offering security guarantees similar
to those of software diversity, control-flow integrity, and information hiding.
We present \emph{\Hydra}, a scheme that deploys a new technique for in-place
code encryption to hide the code layout of a randomized binary, and restricts
the control flow to a benign execution path. This allows us to i)~implicitly
restrict control-flow targets to basic block entries \emph{without} requiring
the extraction of a control-flow graph, ii)~achieve \textit{execution
integrity} within legitimate basic blocks, and iii)~hide the underlying code
layout under malicious read access to the program. Our analysis demonstrates
that \Hydra~is capable of preventing state-of-the-art attacks such as
just-in-time return-oriented programming (JIT-ROP) and crash-resistant oriented
programming (CROP). We extensively evaluate our prototype implementation of Scylla and
show feasible performance overhead.  We also provide details on how this
overhead can be significantly reduced with dedicated hardware support.

\end{abstract}

\section{Introduction}

Instruction set randomization (ISR)~\cite{kc2003countering,
barrantes2003randomized,boyd2010general,portokalidis2010fast} is a
countermeasure initially proposed with the objective of preventing
code injection attacks. ISR provides a unique instruction set for every program
by encrypting its underlying instructions at the binary level. This defense is
effective against code injection because only properly encrypted instructions
will execute on an ISR protected system. Rather than executing directly,
injected code would first be decrypted and then executed. In practice,
this results in an illegal instruction sequence being executed that causes the
program to crash.  With the onset of data execution prevention (DEP), however,
code-injection style attacks became less practical. Instead, adversaries shifted 
to code-reuse attacks (CRAs) as their primary method of exploitation. 

Rather than executing injected code, CRAs craft malicious computations by
stitching together code chunks (gadgets) already resident in the executable
segments of a process. ISR is not an effective defense against CRAs because its
payload has already been correctly encrypted. Malicious reuse of existing
instructions in an ISR protected application results in the correct decryption
and execution of those instructions. To make matters worse, CRAs are not easily
prevented as recent research~\cite{snow2013just, gawlik2016enabling,
bittau2014hacking, evans2015control, carlini2015control, conti2015losing,
schuster2015counterfeit, goktas2014out, evans2015missing} has demonstrated.
These attacks are capable of bypassing state-of-the-art protections, including
address-space layout randomization~\cite{team2003pax}, Google's IFCC
\cite{lattner2004llvm}, and Microsoft's EMET~\cite{emet}.

Prior defenses aimed at preventing CRAs can be loosely categorized into
software diversity~\cite{giuffrida2012enhanced,crane2015readactor,lu2015aslr},
control-flow integrity~\cite{zhang2013practical,tice2014enforcing}, and
information hiding~\cite{backes2014you,gionta2015hidem, tang2015heisenbyte,
kuznetsov2014code}. Software diversity attempts to prevent the adversary from
learning the code or data layout of a program by randomizing program segments,
such as the base address of shared libraries~\cite{team2003pax}. Control-flow
integrity limits the attacker from arbitrarily manipulating the instruction
pointer by checking that every transfer aligns with a pre-computed
control-flow graph of the program~\cite{abadi2005control}. Code Pointer
Integrity (CPI) and information hiding techniques attempt to prevent leakage of
code or data by isolating sensitive structures, such as code pointers or
executable pages in protected memory regions~\cite{kuznetsov2014code,
crane2015readactor}. However, as we elaborate in Section~\ref{sec:related},
each of these defense strategies have their strengths and weaknesses. Until now,
there exists no proposal that unifies their individual strengths into one
framework to tackle the threat of code-reuse attacks.

\medskip \noindent\textbf{Goals and Contributions. }In this paper, we present
\textit{\Hydra}, a defense that updates ISR to provide measurable protection
against code-reuse attacks. It combines the salient features of software
diversity, control-flow integrity, and information hiding approaches to
reliably protect against adversaries that reuse or disclose the code layout of
a vulnerable application. The core feature of \Hydra~is a new form of ISR that
combines fine-grain code diversification with per-basic-block encryption to
hide the code layout of a randomized binary, while at the same time restricting
control flow to a benign execution path. This allows us to achieve what we call
\textit{execution integrity} within legitimate basic blocks. Execution
integrity offers protection similar to coarse-grain control-flow integrity
without the need to statically or dynamically compute a control-flow graph.
\Hydra~also affords complete read-access to a vulnerable program without
revealing the underlying code, due to per-basic-block encryption, or its layout
due to basic block diversification. To the best of our knowledge, \Hydra~is the
first defense that demonstrates how ISR can be extended to protect against
code-reuse attacks. 

In summary, we make the following contributions: 
\begin{itemize} 
	\itemsep0em
    \item \textbf{Per-Basic-Block ISR}: we present a new form of ISR that
    offers protection against CRAs and prevents disclosure of the underlying
    code layout, including its information and control flow, under complete
    read-access to the program.  
    \item \textbf{Execution Integrity}: we present a novel and practical
    protection technique to prevent an adversary from hijacking benign control
    flow.  
    \item \textbf{Prototype Implementation}: we provide a fully-working 
    prototype implementation for x86\_64 systems that is also capable of
    handling shared libraries.  
    \item \textbf{Extensive Evaluation:} we provide an extensive security and
        performance evaluation. We show that \Hydra is resilient to traditional
        ROP, JIT-ROP~\cite{snow2013just},
        CFB~\cite{carlini2015control,evans2015control}, and
        CROP~\cite{gawlik2016enabling}. Our performance measurements show that
        our system only incurs 20\% overhead, despite the
        complexity of \Hydra's hybrid defense. We also discuss hardware extensions in
        Section~\ref{sec:performance_eval} to further reduce the overhead.
\end{itemize}

\section{Instruction Set Randomization} \label{sec:isr-related}
Instruction Set Randomization (ISR) was initially proposed by Gaurav et al
\cite{kc2003countering} with the objective of countering code-injection
attacks. A static key is utilized to encrypt the entire binary, which is
stored as part of the header of the executable file and loaded into the kernel.
Subsequent accesses to the key are done by the operating system only.

To account for possible deficiencies in utilizing a single key per cipher
block, another ISR approach was proposed by Barrantes et al in
\cite{barrantes2003randomized} that utilizes a one time pad (OTP) to encrypt
memory.  The OTP key is generated by a random number generator.  Although the
platform can be exploited by disclosing the unprotected key file, the work
presented by Sovarel et al in \cite{sovarel2005s} demonstrates a stronger
attack by showing that it is only necessary to obtain the key for some parts of
the memory to compromise the entire system.

The system in~\cite{barrantes2003randomized} was extended by Portokalidis et al
in~\cite{portokalidis2010fast} to add support for dynamic libraries, key
management, and forgoes the utilization of an emulator, opting instead to use
Intel's PIN tool. Memory protection is added for writes to avoid an attacker
overriding the callbacks used by PIN in order to leak information or obtain
arbitrary code execution in the system. Memory reads are not protected,
allowing for encrypted page leakage and key leakage at runtime.

Since previous software based approaches exhibited large performance overhead
a hardware-supported solution, called ASIST~\cite{papadogiannakis2013asist},
was presented to overcome this. They demonstrated
a significant reduction in performance, only 1.5\% on average, with minimal
additional hardware.

\medskip
\noindent\textbf{Updating ISR for Code-Reuse Attacks.}
We note that past ISR defenses were designed with
the prevention of code injection attacks in mind. They do not prevent code-reuse 
attacks as control flow can be redirected to any encrypted instruction
and it will be decrypted correctly. In this work we improve upon ISR by
encrypting a program in basic blocks, which allows us to implicitly enforce
coarse-grained control-flow integrity.

\begin{figure*}[!tbh]
    \centering
    \includegraphics[width=0.95\textwidth]{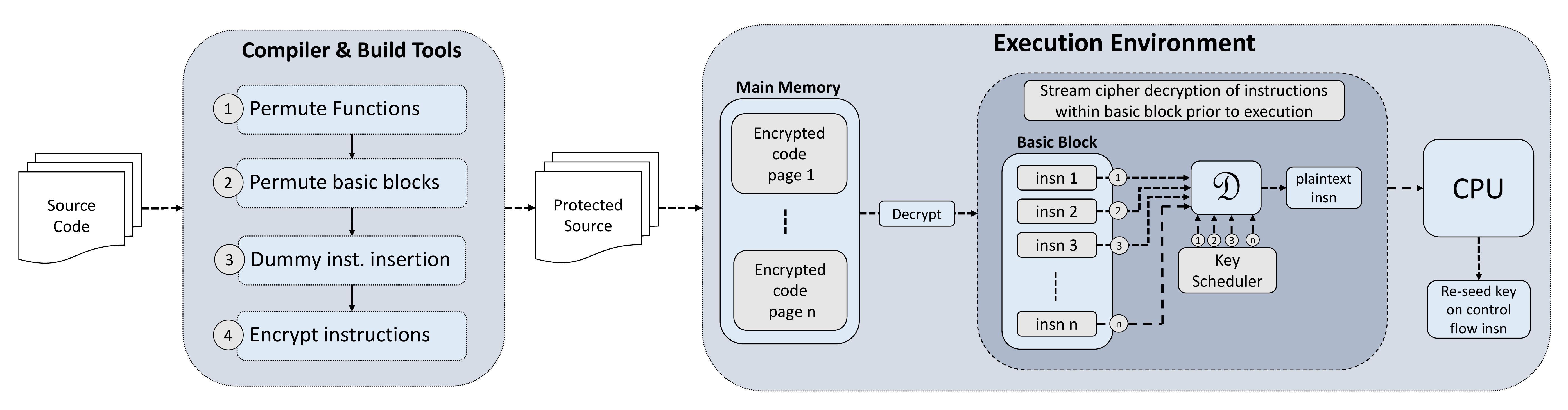}
    \caption{\footnotesize System overview. Our compiler and build-tools
    generate a diversified application using function and basic block
    permutation, as well as dummy instruction insertion. The code is then
    encrypted using a stream cipher over basic blocks. In the execution
    environment, code pages are always encrypted in memory. Basic blocks
    are decrypted live before being forwarded to the CPU.}
    \label{fig:system}
	\vspace{-0.2in}
\end{figure*}

\section{Threat Model}\label{sec:threatmodel}
We assume an adversary equipped with a memory corruption vulnerability, which
allows arbitrary read operations to any mapped memory page of the process.  We
also allow the attacker to perform arbitrary writes to the data space of the
process. We do not allow the modification or injection of code in memory as we
assume code to be protected through DEP~\cite{andersen2004data}. We also assume the attacker
has access to an unprotected binary as a reference, but does not know the 
in-memory code layout of the executing victim process partly due to ASLR~\cite{team2003pax}, 
but also because of \Hydra's protection.

We assume that the attacker's goal is to statically or dynamically reveal code
pages to discover the code layout and craft a code-reuse attack. The attacker
can follow code pointer references in the application's data memory, dynamic
linking segments such as the global offset table (GOT) or through the dynamic disassembly of code
pages. We consider a successful attack against \Hydra~ to be one that reliably
discloses the diversification secrets. The attacker can then use the secrets to
hijack the program's control flow for malicious purposes, such as disabling
DEP to execute injected code. We do not consider
data-flow based attacks, wherein the attacker modifies key pieces of data in
memory to obtain an alternate but legal control flow. 

To summarize, our threat model is in line with previous
research on ISR, but additionally considers the important class of code-reuse
attacks as its main defense target.

\section{\Hydra~Design}\label{sec:hydra_design}
\Hydra~extends Instruction Set Randomization (ISR) to protect against
code-reuse attacks. Classic ISR was designed with code injection attacks in
mind. The design does not readily translate to protecting against code-reuse
attacks (CRA), since instructions under classic ISR are decrypted correctly
regardless of their position in program code. \Hydra, therefore, incorporates
the notion of an execution path within a basic block.\footnote{We consider a
basic block to be any sequential code sequence with a unique entry and unique
exit and use this definition throughout the paper.} In particular, each
instruction within a basic block is decrypted with respect to its predecessors.
This guarantees sequential execution from a basic block's unique entry to its
unique exit. We call this protection \emph{execution integrity}. \Hydra~further
addresses known plain-text to cipher-text attacks, of which past ISR approaches
were vulnerable~\cite{sovarel2005s}, by diversifying the code layout.

\Hydra~protects applications from code layout disclosure, recovery of the
program's underlying information and control flow, and control flow hijacking.
\Hydra~applies fine-grained code diversification by permuting
functions~\cite{kil2006address} and basic blocks~\cite{wartell2012binary}
within functions, as well as inserting dummy
instructions~\cite{jackson2013diversifying} to prevent disclosure of the
program's layout in memory. To hide the program's underlying information and
control flow, \Hydra~encrypts all code pages utilizing a stream cipher over
every basic block in the program. Instructions are decrypted as they are
scheduled to be executed, closing the threat of dynamic disassembly via an
attacker controlled arbitrary read. Implicit control-flow integrity, or
execution integrity, is supported by implementing per-basic-block encryption
which constrains control flow to function and basic block entries.

Figure~\ref{fig:system} shows the major design components of \Hydra. The
program is first diversified by permuting the locations of functions in the
code segment. Then, basic blocks inside functions are permuted and dummy
instructions inserted into those basic blocks. After diversifying the
application, we encrypt every basic block using a stream cipher. The execution
environment ensures that the protected application is always encrypted in
memory. Encrypted instructions scheduled for execution are decrypted using the
stream cipher as they enter the CPU. Plain-text instructions are forwarded to
the CPU and execute normally. Decrypted instructions are never written back to
memory and read access to code pages return cipher-text. We describe each
component in detail below.

\medskip
\noindent\textbf{Fine Grained Code Randomization:} \Hydra combines several diversification
techniques to complement in-place code encryption. In the following we explain
how we combine function permutation, basic block reordering, and the insertion of
dummy instructions to diversify the binary layout of the protected program.
While all of these techniques have been previously proposed~\cite{giuffrida2012enhanced,
wartell2012binary, onarlioglu2010g}, combining them is
challenging and has not been done before. 

First, we permute function locations to hinder disclosure attacks that reveal the
program's layout by reading code pointers in data pages. Usually, an adversary with
arbitrary read capabilities is able to correlate code pointers with target instructions.
By leveraging function permution, we reduce this capability of the attacker to
guessing.

Second, we permute basic blocks within functions to conceal the function
layout. This has two effects: on the one hand, it increases the complexity of our
diversification by adding another layer of randomization.
On the other hand, this protects the encrypted program from known plain-text attacks.
For instance, if a function entry is discovered the attacker
can correlate the encrypted instructions against her local copy to compute the
key.  However, this cannot be achieved without knowledge of the basic block
ordering.

Third, we randomly distribute dummy instructions within basic blocks, altering
address offsets and instruction ordering within a basic block. Dummy
instruction insertion also complements \Hydra's encryption mechanism by
concatenating a random string to the program.  In this case, the type (i.e.,
\texttt{mov, add, push/pop, nop}), number, and distribution of inserted instructions
within a basic block make-up the random string.

The security guarantees offered by \Hydra~are not tied to diversification alone,
but to the combination of randomization with per-basic-block encryption.
Disclosure of a single, or several, function or basic block entries
does not reveal the surrounding code layout when the program is diversified.
However, an adversary can recover portions of the code layout, or even the
whole code layout, by exploiting the techniques described in
Section~\ref{sec:related}. In this case, per-basic-block code encryption
complements diversification by ensuring that the control flow of the program
can only target valid basic block entries and that the code residing at the
disclosed addresses remains concealed as cipher-text.

\medskip
\noindent\textbf{Per-Basic-Block Instruction Encryption:} After randomizing the
binary, we encrypt the program to hide the underlying code layout and enforce
execution integrity.  The purpose of this additional protection is to force the
control flow to the unique entry and exit points of a basic block to prevent
control flow hijacking.  In particular, \Hydra encrypts every basic block in
the program using a stream cipher.  This guarantees that in order to correctly
decrypt and execute the application, control flow must target the unique entry
point of a basic block and execute sequentially to its terminating control flow
instruction.  This form of implicit control flow integrity due to \Hydra's
encryption method is what we have called \textit{execution integrity}.
\Hydra's encryption combined with diversification also protects against
disclosure because reading the code location does not reveal any other
information about the surrounding code layout.  In fact, an adversary would
need to break the underlying encryption to reveal the program's in-memory
layout, as directly reading code pages only returns cipher-text. 

\medskip
\noindent\textbf{Runtime Decryption:}
As depicted in Figure~\ref{fig:system}, instructions are decrypted as they are
fetched by the CPU. This ensures that no instruction is ever in plain-text in
main memory or any of level of the shared cache hierarchy. When a control flow instruction
is encountered, the execution environment reseeds the stream cipher for the
next basic block. \Hydra's execution environment guarantees that the underlying
code layout, including its information and control flow, remains hidden from
an adversary. Hijacking control flow must occur along existing control flow
paths within the program in order to decrypt instructions correctly.

\section{Architecture}\label{sec:hydra_arch}
In the following, we describe our architecture for \Hydra's fine-grained
randomization and code encryption mechanism in detail.  To randomize the
binary, we modify the LLVM compiler infrastructure~\cite{lattner2004llvm} to
generate a diversified program that supports per-basic-block instruction
encryption by permuting functions, reordering basic blocks within functions,
and randomly distributing dummy instructions within basic blocks.  We note that
source code is not a strong requirement for randomization, and that similar
diversification techniques have already been applied at the binary
level~\cite{wartell2012binary,davi2013gadge}. However, for our proof-of-concept
we chose to leverage the extensive functionality of LLVM for program
transformation and ease of implmementation.  For \Hydra's encryption mechanism,
we extend standard tools for binary manipulation on Linux.

\medskip \noindent\textbf{Fine Grained Code Randomization:} We permute
functions, and then basic blocks within functions, to improve the cache
pressure that might otherwise be incurred if basic blocks were randomly
distributed across the entire executable. Although this reduces the entropy of
the randomization, we reason that this decision does not sacrifice security. In
Section~\ref{sec:security} we demonstrate that the number of possible
permutations using the SPEC CPU2006 benchmark suite~\cite{henning2006spec} is
sufficiently high to prevent an adversary from revealing the code layout.

We randomly distribute dummy instructions within
basic blocks in the program by selecting from \texttt{nop} instructions,
instruction that move a register onto itself, arithmetic and logic instruction
with an identity element, and pushing and popping a register from the stack.
We include this diversification technique to prevent cipher-text only and known
plain-text attacks to which past ISR approaches have been
susceptible~\cite{sovarel2005s}. Launching such an attack against \Hydra~after
dummy instruction insertion requires an adversary to know the number,
distribution, and type of dummy instructions inserted within the basic block.

\medskip \noindent\textbf{Encrypting Code Pages:} \textbf{Creating a
\Hydra~binary:} Encryption in our \Hydra~prototype uses an extended
\texttt{objcopy} program, part of the \texttt{GNU binutils} toolset.  Normally,
\texttt{objcopy} is used to copy and translate object files between different
formats. We extend its functionality to output encrypted executables using a
stream cipher.  We gate this functionality with a command line option. When
active, code pages are disassembled as they are copied using \texttt{libopcode}
as the backend for disassembly. Basic blocks are then encrypted using a stream
cipher and the encrypted code is written to a new executable.

Our modified \texttt{objcopy} stores the seeds used in the stream cipher and
the cipher permutation in a new section of the binary. This section, named
\texttt{.encseeds}, is flagged as non-allocatable, which means that the loader
will not store it in memory.  Our execution environment (see
Section~\ref{sec:runtime_wrapper}) utilizes the information within this section
to decrypt the binary live as it runs.

\Hydra's encryption capabilities facilitate our principle of \emph{execution
integrity} by ensuring that instructions are executed in the way they were
intended. This is achieved by encrypting every instruction with respect to its
predecessor using a stream cipher; the first instruction in a basic block chain
is encrypted with respect to a seed.  Previous code encryption approaches
encrypt code pages regardless of instruction sequences and were therefore
unable to protect against code reuse attacks.

\Hydra's encryption model applies to standalone, well-formed basic blocks. This
is deliberate in order to avoid whole program analysis and recovery of a
program's complete execution path in order to correctly manage encryption keys.
Instead, \Hydra~circumvents this and simplifies key management by using a
single key to encrypt the first instruction at a control flow target. This
simplification has the side-effect of allowing \Hydra~to implicitly handle
problematic code constructs, including calls to library functions and signal
handlers, e.g., by using key initialization on basic block entry. \Hydra does
not currently support JIT code generation.  

Each encryption chain is terminated when a control flow instruction is
encountered. For alignment purposes, compilers may issue dummy instructions as
padding at the end of a function. As such, the entry point of an ensuing
function may not immediately follow a control flow instruction. To avoid missed
basic blocks during the encryption process, we have \texttt{objcopy} read the
symbol table of the executable and look for function entries, and encrypt them
to ensure correctness in the process.

\noindent\textit{Key Generation:} For our prototype implementation, we utilize
a pseudo-random number generator (PRNG) to generate a key stream for our stream
cypher. We utilize the Advanced Encryption Standard in Counter Mode (AES-CM) to
generate keys. Since the largest acceptable x86\_64 instruction at the time of
writing is 15 bytes wide, we utilize AES-CM with a 128 bit block size and
ignore the uppermost byte for the encryption of instructions. AES-CM proves to
be sufficiently resilient for our purposes.  We seed the AES-CM engine with a
key and initialization vector on every basic block entry. Our counter function
is a simple increment by one.

\medskip
\noindent\textbf{Runtime Wrapper:}\label{sec:runtime_wrapper}
To the best of our knowledge, there is no commercially available platform
capable of decrypting instructions as they are executed. In prior
instruction set randomization schemes \cite{kc2003countering,
barrantes2003randomized, boyd2010general, portokalidis2010fast} dynamic binary
instrumentation tools, such as Dynamic RIO or Intel PIN, have been used to
perform live decryption. However, Dynamic RIO and Intel PIN share their own
address space with the binary they are instrumenting. Under our attacker model,
this means that the diversification and encryption secrets can be directly
leaked using memory disclosures. Although these disclosures can be prevented by
checking the locations from where a memory read is taking place, it would
not be a realistic emulation of the necessary hardware to accomplish the task
of runtime decryption.

System emulators such as QEMU or MARSSx86 are possible alternatives to the
implementation, but come with their own set of challenges. These tools act as
hypervisors and as such come with the requirement of a guest operating system.
Coupled with this requirement is the need to handle multitasking and
applications that do not support our scheme. As such, a great deal of
modification would be required for a testing platform based on these tools.
QEMU is capable of running applications emulating a full system's user land
without the requirement of a guest operating system. However, when used this
way, it translates the guest program's instructions to native code and caches
it in a set of translation blocks. Although this technique is used to improve
runtime performance, it would be in direct violation of our requirement of
keeping all code pages encrypted in memory.

Consequently, we developed a prototype simulator for \Hydra which closely
resembles our proposed hardware model. The simulator consists of three major
components: a Launcher, Monitor, and Encrypted Process. When invoking the
simulator, the Launcher process runs first. The Launcher reads the contents of
the \texttt{.encseeds} section and obtains the seeds for the stream cipher and
the permutation applied to it.  At this point, the Launcher spawns a child
process which requests to be traced using the \texttt{ptrace()} facilities of
the Linux kernel. The child then spawns the Encrypted Process and waits for the
parent, which has been converted to a Monitor. The Monitor process proceeds to
trace the execution of the Encrypted Process until the latter terminates.

Algorithm~\ref{alg:parent_monitor_proc} delineates the operation of the Monitor
process. As long as the Encrypted Process is in a runnable state, the Monitor
obtains the program counter of the child and fetches 15 bytes from that address
into a local buffer. If the last instruction executed by the Encrypted Process
was a control flow instruction, the Monitor reseeds the stream cipher with
information obtained from the Loader process, otherwise, the Monitor forwards
the stream cipher to the next state. The data stored in the local buffer is
then decrypted using a key obtained from the stream cipher. The Monitor
decodes the instruction while keeping record of the instruction type, and writes
the decrypted instruction to the Encrypted Process' code segment at the location
of the program counter and singlesteps it. This allows the Encrypted Process to
execute one instruction at a time. The Monitor then writes back the encrypted
buffer into the Encrypted Process as it would otherwise leave portions of code
as plain-text in memory.

\begin{algorithm}
	\caption{Monitor process operation. The Monitor Process singlesteps through
	the Encrypted Process decrypting instructions as they are about to be
	executed.}
	\label{alg:parent_monitor_proc}
	\begin{algorithmic}[1]
		\While{child is running}
		\State{$pc \gets child_{\%rip}$}
		\State{$buf \gets child_\texttt{.text @ pc}$}
		\If{last isns is cflow}
			\State{Reseed stream cipher}
		\Else
			\State{Forward stream cipher}
		\EndIf
		\State{$buf \gets buf \oplus key$}
		\State{Decode instruction}
		\State{$child_\texttt{.text @ pc} \gets$ insn}
		\State{Singlestep child}
		\State{$buf \gets buf \oplus key$}
		\State{$child_\texttt{.text @ pc} \gets$ buf}
	\EndWhile
	\end{algorithmic}
\end{algorithm}

Our simulator closely resembles the operations a  hardware-based implementation
would take. In a Linux-based system, the kernel reads the sections of an
executable and allocates them in memory as needed. The kernel can be extended
to perform the tasks of our Loader process. Load-time encryption can be
achieved by extending the kernel loader. The Monitor process simulates a
combination of a hardware decryption unit and kernel process manager. As
instructions are fetched, they are decrypted using the stream cipher by a
hardware module in the front-end. This hardware module handles, in conjunction
with the kernel, context switches and multitasking. Much like in our execution
wrapper model, the Encrypted Process is unable to leak secrets from the Monitor
because a hardware-based model keeps the secrets by enforcing memory
separation.

Since our simulator runs as a user space process, we provide support for
dynamic linking by exploiting the workings of the dynamic loader in UNIX-based
systems. Before launching the encrypted application, we set the
\texttt{LD\_LIBRARY\-\_PATH} environment variable in the child process and
point it to a path where the encrypted libraries reside. The child application
also requests an encrypted resolver. A library function call proceeds in much
the same fashion as in a normal system. The resolver gives priority to the
encrypted libraries located in the directory specified by
\texttt{LD\_LIBRARY\_PATH}. The only information that is needed are the seeds
used by the stream cipher to decrypt the library and loader information. This
is obtained from the \texttt{.encseeds} section of the object files. We then
tie the seeds to the address range of the library and allow execution to
proceed as normal. As such, encrypted libraries can be freely shared across
applications that use them in a rich operating system environment.

The simulator also allows us to monitor other aspects of the child process, such
as the time that was spent during execution, and the stream of instructions
being executed. We profile a set of applications from the SPEC CPU2006 benchmark
and discuss results in Section \ref{sec:performance}.

\section{Security Evaluation} \label{sec:security}
We perform our security evaluation assuming that the attacker's goal is to
disassemble enough code pages, or infer enough code locations, to craft a
code-reuse attack (CRA). As such, we analyze and evaluate the effectiveness of
our scheme in relation to an adversary equipped with a memory vulnerability
that allows disclosure of the program's address space and the ability to
arbitrarily redirect control flow of a program to any chosen target. This is in
line with recent work~\cite{wartell2012binary,hiser2012ilr,kuznetsov2014code,
crane2015readactor, crane2015trap}.  Following standard cryptographic
protocols, we assume that the attacker knows the workings of the underlying
implementation of our stream cipher, but does not know the keys and IVs used to
seed the pseudorandom number generator.

\medskip \noindent\textbf{Code-Reuse Attack Payload:} As mentioned in
Section~\ref{sec:threatmodel}, we assume that the attacker cannot modify or
inject code in memory due to data execution prevention (DEP). Because of this,
almost all real-world CRAs aim at disabling DEP and launch a code-injection
attack thereafter.  Particularly, we assume an adversary, who is equipped with
a memory-disclosure vulnerability.  Usually, this allows bypassing even
fine-grain diversification and chaining function or basic block gadgets to
construct an attack that disables DEP.  However, to bypass \Hydra, the attacker
would still need to break the per-basic-block encryption to perform useful
operations within the process.  \Hydra~provides built-in protection in this
scenario because it always expects to decrypt encrypted instructions. This
holds despite the permissions of the process.

\medskip \noindent\textbf{Traditional ROP:} Traditional ROP utilizes known
gadgets in the program's code and chains them either using a dispatcher gadget,
as in the case of jump-oriented programming~\cite{checkoway2010return,jop}, or
by corrupting return addresses found on the stack~\cite{Sh2007ds}. Gadgets are
any sequence of instructions found within basic blocks that end with an
indirect branch instruction. Because the x86\_64 instruction set allows
unaligned instructions, gadgets need not be made of intended instructions but
can be made from partially decoded instructions.

\Hydra~protects against control flow redirection to arbitrary instructions
inside a basic block. This is achieved by encrypting basic blocks with a stream
cipher until the terminating control flow instruction on the basic block is
found. During encryption, as new instructions are encountered in the basic
block, the key scheduler is forwarded and a new key is utilized to encrypt the
instruction. During execution, when a new basic block is entered, instructions
are decrypted in a similar fashion. Redirecting control flow to the middle of a
basic block results in instructions being decrypted with the wrong key. This
results in the CPU executing instructions in a pseudostochastic fashion that
eventually causes the program to crash. Previous analysis demonstrates that an
average of four to five instructions is sufficient to terminate a
process~\cite{barrantes2003randomized}.

An attacker who intends to perform a traditional ROP attack would need to
reverse the stream cipher in order for malicious control flow redirection to
work.  With our usage of AES-CM, the attacker must know the secret key and
initialization vector used to seed the AES engine in order to deduce the keys
for decryption.  Our usage of AES-CM ensures sufficiently secure pseudorandom
numbers to encrypt a program in memory.

\medskip \noindent\textbf{Control-Flow Bending:} Control-flow
Bending~\cite{carlini2015control} (CFB) is a recent attack that bypasses CFI
protection assuming an ideal CFI policy is in place with relaxations in the
enforced control flow graph. In a CFB attack, an adversary corrupts a code
pointer to call a valid function entry, as determined by the CFI policy. The
callee must also contain a vulnerability which allows the attacker to corrupt
the return address, pointing to any call-preceded site.  This allows the
attacker to bend control flow maliciously and craft any arbitrary exploit.
However, \Hydra~employs fine-grained diversification with per-basic-block code
encryption to prevent the attacker from locating call-preceded targets. This
attack further requires the adversary to locate a vulnerable function in the
program allowing her to overwrite a return address. As discussed, our
protection efficiently prevents knowledge of the underlying information and
control flow of the program as well as its code layout due to the combination
of fine-grain code diversification with per-basic block encryption.

\medskip \noindent\textbf{JIT-ROP:} JIT-ROP~\cite{snow2013just} is a powerful
class of attacks against fine-grained randomization techniques. In particular,
these attacks exploit the disclosure of a single code pointer to adjust their
ROP payload to the randomized program layout and gadget search space at
runtime. This enables an adversary to also defeat load-time based defense
approaches that re-randomize the application program space layout per
execution. We discuss the two main types of JIT-ROP and how they are handled by
\Hydra.

Conventional JIT-ROP~\cite{snow2013just} disassembles code pages while keeping
a collection of gadgets found during disassembly. Any code pointer, as part of
a direct call or a direct jump, is used to find new code pages, bypassing the
diversification applied to the program. However, under \Hydra, any read
performed from a code page yields cipher-text. Although the code page itself is
readable, it cannot be readily disassembled without being able to break the
encryption. As such, an attacker is unable to utilize conventional JIT-ROP to
disclose new code pages or find gadgets in the binary.

JIT-ROP can be extended to use only code pointers disclosed from data pages, or
indirect disclosure, as the pivot point to start a disassembly chain of code
pages~\cite{davi2015isomeron}. Although \Hydra~does not fully protect against
indirect disclosure, it does prevent disclosure of the entire program due to
the underlying encryption. \Hydra's protection prevents such a vulnerability
from revealing the surrounding code layout because of fine-grain code
diversification combined with per-basic-block encryption.

The indirect disclosure of function entries or basic blocks provides the
attacker with potential gadget locations inside basic blocks. To tackle this
type of disclosure, other approaches such as
Readactor~\cite{crane2015readactor} utilize trampolines to hide the location of
basic blocks, eliminating any indirect disclosures of gadgets inside them.
\Hydra, on the other hand, prevents the usage of those gadgets via execution
integrity. Arbitrary redirection of control flow to gadgets within a basic
block results in the execution of incorrectly decrpyted instructions, causing
the program to crash.  As such, \Hydra~is able to forgo the usage of
trampolines and the separation of code from data. Furthermore, an attacker who
discloses a basic block is faced with the same problem encountered when
attempting to disclose new code pages with conventional JIT-ROP, breaking the
encryption. We show that this requires knowledge of the secrets behind the key
scheduler, which proves to be infeasible as described in
Section~\ref{sec:diversification_eval}.

For future work, we plan to extend our protection mechanism to better conceal
code pointers in memory to avoid indirect memory disclosure.  \Hydra~could be
extended to encrypt vtable pointers and vtables and decrypting them as they are
about to be used.  Futhermore, we will investigate novel ways to perform load
time randomization of binaries and more efficient methods to encrypt
instructions in memory.

\medskip \noindent\textbf{Crash-Resistant Oriented Programming:} Crash
Resistant Oriented Programming (CROP)~\cite{gawlik2016enabling} is a powerful
attack capable of bypassing fine-grained randomization, information hiding, and
control-flow integrity protection on client-side applications. This attack
exploits mishandled exception handling and system call behavior to scan memory
without crashing, transforming these crash resistant primitives into so called
\textit{memory oracles} allowing the adversary to infer accessible memory
boundaries. Combining these two techniques, the authors demonstrate the ability
to locate unreachable memory regions such as the thread and process environment
block in Windows systems, reference{-}less memory regions such as the safe
regions in code-pointer integrity (CPI)~\cite{kuznetsov2014code}, subvert
hidden code pages by locating export symbols or trampolines addresses used in
Readactor~\cite{crane2015readactor}, and discover functions or valid return
targets within a control flow path.

Although \Hydra~does not prevent an attacker from reading code pages using a
memory oracle, any reads performed from code pages return cipher-text. Due to
both the key permutation and code diversification techniques offered by our
protection, correlating plain-text to cipher-text is infeasible. It could be
reasoned that an attacker could use a crash-resistant memory oracle to redirect
control flow into the cipher-text to locate valid control flow targets.
However, the memory oracles used in CROP are only resistant to crashes from
segmentation faults. Memory oracles do not handle bus errors or illegal
instruction faults, which are the main causes of program termination under
\Hydra's protection when unintended instruction sequences are executed.

\medskip \noindent\textbf{Comparison with Binary CFI:} Our goal is to protect
an application against sophisticated code-reuse attacks through an
instrumentation on the \emph{binary} level. This means that \Hydra~does not
require the source code of the protected program and is applicable to a large
range of software. While for our proof-of-concept solution we chose to
diversify programs using the LLVM compiler for ease of implementation, source
code is not a strong requirement~\cite{wartell2012binary, davi2013gadge}.
Furthermore, even though control-flow integrity (CFI)~\cite{abadi2005control}
was initially proposed as a compiler extension, this has been extended to the
binary level as well~\cite{zhang2013control,zhang2013practical}.  However,
these extensions have been shown to be vulnerable to attacks because of their
relaxed CFI policies~\cite{goktas2014out,davi2014stitching,carlini2014rop}.
These policies are due to the inherent difficulty of reconstructing an accurate
control flow graph (CFG) of a program from its binary representation.

Our approach improves upon the inherent limitations of binary CFI schemes,
because \Hydra~does not require a CFG of the protected program as input.
Instead, we protect the application through a combination of per-basic-block
encryption and randomization to restrict program execution to the intended path
at runtime.  More importantly, our security guarantees do not depend on the
asserted precision of such a graph. 

Similar to other defenses deployed on the binary level, we do not cover attacks
such as counterfeit object-oriented programming
(COOP)~\cite{schuster2015counterfeit}, or whole-function reuse at the moment.
However, we are currently investigating the possibilities of offering a
protection against these kinds of attacks within our scheme with added vtable
and function protection through trampolines, as proposed by
Readactor++~\cite{crane2015trap}.

\section{Analysis of Encryption and Diversification}\label{sec:diversification_eval}
\medskip \noindent\textbf{Probability of Guessing Basic Blocks:} Encrypting
basic blocks using a stream cipher provides an adversary with the possibility
of maliciously chaining the execution of basic blocks. To maliciously chain
execution, an attacker must disclose the location of the necessary basic
blocks. Thus, we focus on the probability of an attacker locating a chosen
basic block.

\Hydra~randomizes the layout of a binary and then applies encryption over basic
blocks, which provides protection against direct disclosure of code.  An
attacker reading from code is only able to obtain cipher-text and has no
knowledge of the underlying instruction stream because of the randomization.
For our prototype, we ensure that the layout of the resulting binary differs
substantially throughout different compilations. Because we randomize function
locations and reorder basic blocks within a given binary, the amount of entropy
directly depends on the number of functions and basic blocks in a binary.

We reason that by performing direct disclosures of code pages, an attacker
would theoretically be able to locate an individual basic block with a
probability of $1/\sum^m_{i=1} n_i$ where $m$ is the number of functions in a
program and $n_i$ the number of basic blocks in function $i$. Load-time
randomization for a process of $m$ functions and $n_i$ basic blocks in function
$i$ provides the maximum theoretical entropy for our randomization scheme,
given by $\left(\sum^m_{i=1} n_i\right)!$. However, cache locality is lost
during execution and a higher penalty is paid in performance because of the
distribution of basic blocks under this randomization scheme. We find that this
is unneccessary, as we are still able to obtain a large amount of entropy
without significantly degrading cache performance by randomizing function
locations and positions of basic blocks within them. The entropy provided under
this scheme is given by $m! \times \prod^m_{i=1} n_i!$.

\begin{figure}[ht!]
    \centering
    \begin{tikzpicture}
	\begin{semilogxaxis}[
			xlabel={Binary Size},
			ylabel={Probability},
			height=2in,
			width=\columnwidth,
			nodes near coords,
			ymin = -0.0003,
			ymax = 0.0022,
			visualization depends on={\thisrow{alignment} \as \alignment},
			every node near coord/.style={font=\footnotesize,anchor=\alignment},
		];

		\addplot+[scatter, only marks, scatter src=explicit symbolic] table[meta=bmk] {
			x			y				alignment	bmkno	bmk
			63488		0.00191659		90		1		mcf
			83968		0.000869565		-90		2		libquantum
			56320		0.000625978		90		3		astar
			906240		0.000381869		-90		4		omnetpp
			96256		0.000350877		90		5		bzip2
			186368		0.000174484		90		6		sjeng
			356352		0.000085112		-90		7		hmmer
			623616		0.0000479738	90		8		h264ref
			3984588.8	0.0000320407	-90		9		gobmk
			8598323.2	0.0000294829	90		10		xalan
			3460300.8	0.00000597693	90		11		gcc
		};
	\end{semilogxaxis}
\end{tikzpicture}
	\vspace{-0.2in}
	\caption{\footnotesize Probability of disclosing function and basic block
    entries.}
    \label{fig:probs}
	\vspace{-0.15in}
\end{figure}
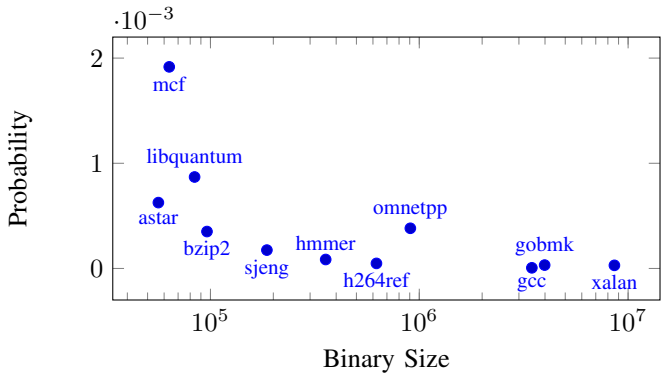

Therefore, in practice, the probability of an adversary locating an individual
basic block through direct disclosure can be estimated as $1/(n_{av} \times
m)$, where $m$ is the number of functions, and $n_{av}$ is the average number
of basic blocks per functions. Using data obtained from the SPEC CPU2006
benchmarks, Table~\ref{tab:stats}, we compute the probability of finding a
basic block and correlate it to the size of the code space of the benchmarks.
This is shown in Figure \ref{fig:probs}. As expected, the probability of
guessing a particular basic block decreases as the size of the code segment
increases.  Generally speaking, larger binaries will contain a higher number of
basic blocks. An attacker exploiting a large binary with a small number of
basic blocks will be faced with basic blocks that have large side effects,
which is detrimental to a CRA.

\begin{table}[h]
     \centering
     \begin{tabularx}{\columnwidth}{X|c|c|c}
	 	\hline\hline
		\textbf{Name}  & \textbf{Mean \#BB/FN} & \textbf{\#FN} &
		\textbf{Size in Bytes}\\
		\hline\hline
		astar & 7.5 & 213 & 55K\\
		\hline
		bzip2 & 28.5 & 100 & 94K\\
		\hline
		gcc & 30 & 5577 & 3.3M\\
		\hline
		gobmk & 11.65 & 2679 & 3.8M\\
		\hline
		h264ref & 35.33 & 590 & 609K\\
		\hline
		hmmer & 21 & 538 & 348K\\
		\hline
		libquant & 10 & 115 & 82K\\
		\hline
		mcf & 21.74 & 24 & 62K\\
		\hline
		omnetpp & 4.35 & 602 & 885K\\
		\hline
		sjeng & 39.8 & 144 & 182K\\
		\hline
		xalan & 6 & 5653 & 8.2M\\
	 	\hline\hline
	\end{tabularx}
	\caption{SPEC2006 basic block statistics. \emph{Mean \#BB/FN} denotes the
	average number of basic blocks per function. \emph{\#FN} denotes the number
	of functions in the binary. The number of permutation for function
	shuffling combined with per function basic block randomization is computed
	as $m!\times\prod_{i=1}^{m}n!$, where $m$ is the number of functions and
	$n$ is the number of basic blocks.}
	\label{tab:stats}
\end{table}

\Hydra~also provides a mechanism to keep similar basic blocks from being
identified by inserting dummy instructions within basic blocks, then encrypting
them. This provides an extra layer of obfuscation against direct direct memory
disclosure because the original program has effectively been concatenated with
a random string. With dummy instructions inserted into a program an attacker is
unable to perform frequency analysis in leaked cipher-texts to identify basic
blocks, or perform known plain-text attacks without first knowing their
distribution.

\medskip \noindent\textit{Attacker Obtains A Basic Block} We now discuss the
effects of the attacker being able to obtain a basic block.  Although this is
not possible using direct disclosure of a code pointer, an attacker may still
be able to obtain some manner of basic block information by disclosing code
pointers such as those found in the GOT, vtables or return addresses stored in
the stack.  We perform an analysis of what can be done with this information
under two circumstances: the attacker does not have matching plain-text, and
the attacker has a matching plain-text.

\medskip \noindent\textit{Attacker does not have matching plain-text:} An
attacker who obtains cipher-text, but does not have matching plain-text may
attempt to study the behavior of the program by altering a code pointer to
force execution from an adjacent location. Upon executing this new code path,
the decryption mechanism will return incorrect instructions. The attacker can
then observe any side effects caused by this execution
path~\cite{sovarel2005s}.  Previously reported experimental data shows that the
probability of a meaningful computation under this conditions is small and that
the program will likely crash due to an illegal memory access or illegal
instruction~\cite{barrantes2003randomized}.

Furthermore, this type of attack is only reliable if the program is respawned
with the exact randomization features as before, such as when naively cloning
code and data pages from a parent process without re-encrypting and
re-randomizing them. If the newly loaded code pages use a new seed and key,
then the attacker is unable to perform side-channel analysis under this
scenario.

\begin{figure*}[ht!]
	\centering
	\includegraphics[width=\textwidth]{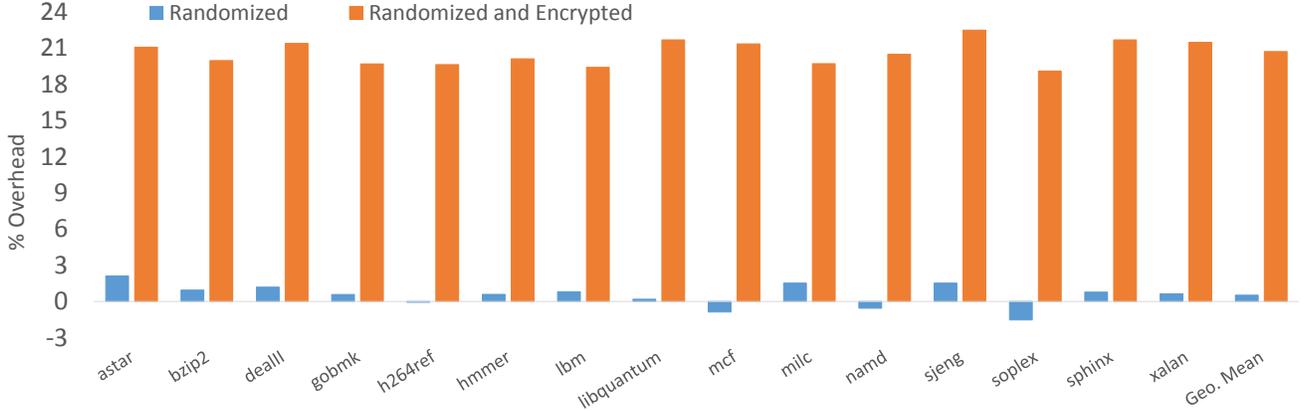}
	\vspace{-0.2in}
	\caption{\footnotesize SPEC CPU2006 performance results.}
	\label{fig:spec2006}
\end{figure*}

\medskip \noindent\textit{Attacker has matching
plain-text}\label{sec:aes_cm_nist} The assumption that an attacker has a
matching plain-text is not unsound. If the disclosed code pointer is a vtable
entry or a GOT entry, for example, then the attacker can disclose the entry
basic block to a function. The attacker can then attempt to utilize a plaintext
copy of the binary to obtain the encryption secrets.

Under these conditions, the attacker is able to obtain $n$ bits of the key by
performing a bitwise XOR between the first $n$ bits of the encyrpted block and
the first plaintext instruction in the basic block. The attacker may then
attempt to find the rest of the keys on a basic block by performing subsequent
decryption operations. Using this information, an attacker can iteratively
attempt to reverse the secrets behind the stream cipher until reaching the
basic block's terminating control flow instruction and obtain its target
address. We randomly distribute a permutation of dummy instructions within
basic blocks to effectively counter this attack by concatenating a random
string to the plain-text and then generating the cipher-text.

If a predictable key scheduler is used the attacker can potentially use this
information to guess the next key given dummy instruction insertion. Dummy
instructions, such as \texttt{nop} sleds or moving a register onto itself, can
be detected by the attacker since the key obtained from performing a bitwise
XOR between the plain-text instruction and cipher-text instruction would not
match the expected key. An attacker can decrypt the entire basic block until
reaching the terminating control flow instruction. At this point, the target
address can be decrypted and a new decryption chain started. This process is,
in effect, an extension to JIT-ROP~\cite{snow2013just} to discover code pages
by decrypting instructions when using a predictive key scheduler. However,
relying on a cryptographically secure pseudorandom number generator for key
generation ensures that this type of attack is infeasible.

\section{PerformanceEvaluation}\label{sec:performance}\label{sec:performance_eval}
To evaluate the performance impact of \Hydra, we use the SPEC CPU2006 benchmark
suite which contains a set of representative CPU-intensive programs seen in
real-world applications. We use Arch Linux with Linux kernel version 4.8.4 for
our evaluation running on an Intel Core i7-2600 CPU clocked at
$3.4\si{\giga\hertz}$, with a $32\si{\kibi\byte}$ L1 I/D cache,
$8\si{\mebi\byte}$ of L3 cache, and $16\si{\giga\byte}$ of RAM. We use \Hydra's
modified LLVM compiler to build SPEC with \texttt{musl-libc} and
\texttt{libc++}, rather than \texttt{glibc} and \texttt{stdlibc++}, because
LLVM does not currently support \texttt{glibc} and \texttt{stdlibc++} library
extensions. We statically link these libraries for our evaluation, but this is
not a requirement as \Hydra~initializes a key on every basic block entry which
includes entry into library functions.  Because of this we were unable to build
the \texttt{perlbench} and \texttt{gcc} benchmarks as these rely on internals
and extensions provided only by \texttt{glibc}.

We evaluated the compatible benchmarks against native execution with both
fine-grain diversification only, and with both diversification and
per-basic-block encryption. Each benchmark is executed within our execution
wrapper and evaluated by measuring the time taken to execute $250$ million
instructions. We limit the instruction count as a way to speed up data
collection, reporting the effective CPI of our protection mechanism as \Hydra's
execution wrapper single-steps through the program's code, decrypting
instructions live. Time measurements are based on CPU cycles by reading the
time-stamp counter register before and after exuction. A summary of our results
are shown in Figure~\ref{fig:spec2006}. Overall, we found that \Hydra incurs an
average overhead of 20\% for SPEC CPU2006.

\begin{figure}[ht!]
    \centering
    \includegraphics[width=0.99\columnwidth]{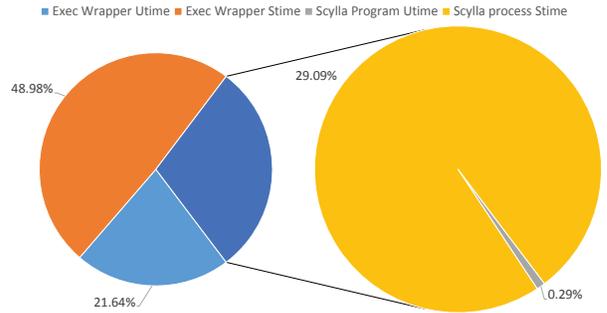}
    \caption{\footnotesize Delineation of the average execution times across all
    SPEC CPU2006 benchmarks with execution wrapper and Scylla protected
    program.}
    \label{fig:spec2006time}
	\vspace{-0.15in}
\end{figure}
\begin{figure}[hb!]
    \centering
    \includegraphics[width=0.99\columnwidth]{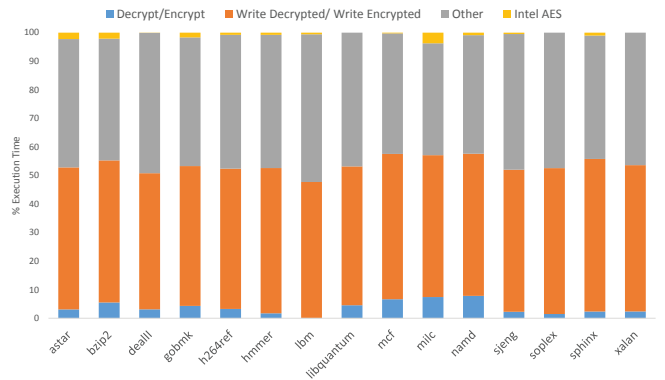}
    \caption{\footnotesize Breakdown of execution times within the wrapper while
    managing a Scylla protected program.}
    \label{fig:execwrappertime}
\end{figure}
\begin{figure*}[ht!]
    \centering
    \includegraphics[width=\textwidth]{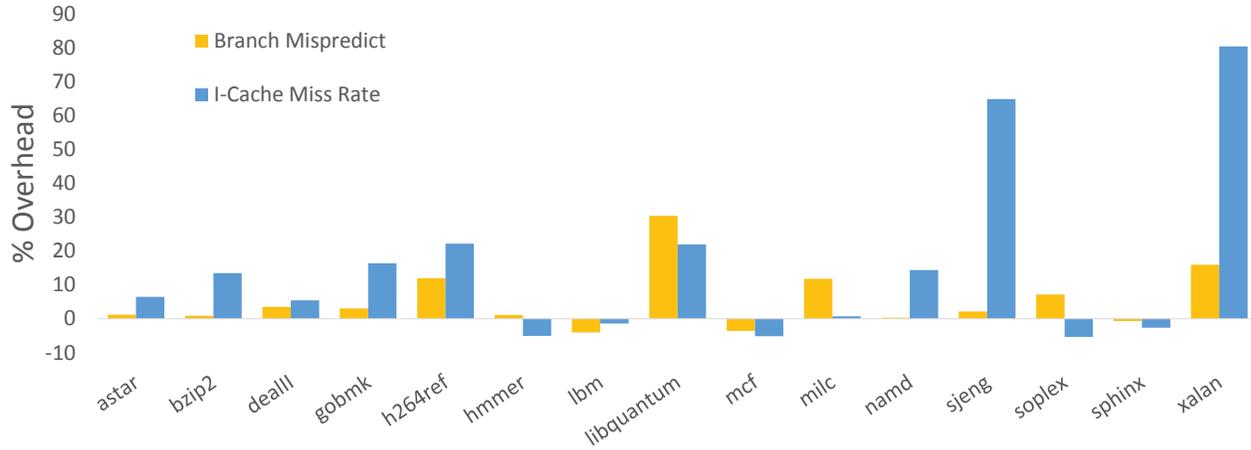}
	\caption{\footnotesize Performance slowdown in Branch Misprediction and
	I-Cache Miss rates due to Scylla's randomization.}
	\vspace{-0.2in}
    \label{fig:perf2006}
\end{figure*}

\medskip \noindent\textbf{Fine Grained Diversification:} First, we evaluate the
impact of code diversification by itself, computing an average overhead less
than 1\%. This performance result agrees with other fine-grain diversification
approaches that randomize basic blocks~\cite{wartell2012binary}. We surmise
that the performance slowdown is due to a combination of imperfect instruction
cache locality because of increased code size as a result of dummy instruction
insertion, or an increase in branch mispredictions due to both basic block and
function permutation.  Figure~\ref{fig:perf2006} shows the performance overhead
caused by \Hydra's diversification for both branch mispredictions and the
instruction cache miss rate.  Branch misprediction is largely low, and in some
cases improved, but on average incurs a 11\% overhead. Both \texttt{libquantum}
and \texttt{xalan} exhibit overheads higher than the average.
\texttt{Libquantum} is a library for simulating a quantum computer and contains
a large number of branches, roughly 26\%, so this result is expected.
\texttt{Xalan} is an XML processor and also contains a large number of branches
compared to the other SPEC CPU2006 benchmarks, so it too experiences an
increase in branch mispredictions.

From our evaluation the instruction cache miss rate suffers the most due to
\Hydra's diversification.  \texttt{Libquantum}, \texttt{sjeng}, and
\texttt{xalan} incur the highest miss rates. Both \texttt{libquantum} and
\texttt{xalan} suffer from poor temporal locality due to basic block and
function permutation combined with a large number of branch instructions. Prior
analysis has demonstrated that the instruction cache performance is sensitive
to both interpreter (\texttt{xalan}), and artificial intelligence
(\texttt{sjeng}) workloads due to large code footprints that execute over a
wide range of functions~\cite{jaleel2010memory}. Overall, \Hydra's
diversification causes a 25\% increase in instruction cache misses across all
SPEC CPU2006 benchmarks.

\medskip \noindent\textbf{Fine Grained Diversification and Instruction
Encryption:} We then evaluated \Hydra's full implementation with both code
diversification and er-basic-block encryption features running within the
execution wrapper. On average, the impact in performance of \Hydra~incurs a
20\% overhead and a worst case overhead of 22.5\%.

\Hydra's~prototype implementation uses a execution wrapper to monitor the
encrypted process and decrypt instructions as they are scheduled to be
executed. This includes reading from the encrypted child process to fetch up to
a 15 byte block, generation of a 128-bit key using Intel's AES-CM engine,
decryption of the instruction, writing the decrypted instruction back into the
child, executing the instruction in single-step mode using \texttt{ptrace()},
re-encrypting the instruction, and then writing the encrypted instruction back
into the child. This is an simulation framework developed to model the security
offered by a hardware-based implementation, which would place the decryption
engine between the instruction cache and front-end of the CPU ensuring that
instructions in the memory hierarchy would always remain encrypted.

Past instruction set encryption approaches have demonstrated that a hardware
supported solution for runtime decryption significantly reduces this
overhead~\cite{papadogiannakis2013asist}. As our protection does not conflict
with this approach, we argue that the overhead shown in
Figure~\ref{fig:spec2006} for both full code diversification and encryption is
misleading in that it is due primarily to the execution wrapper.
Figure~\ref{fig:spec2006time} delineates the average execution times for
\Hydra's full protection. Roughly 71\% of the benchmark execution time is spent
within the execution wrapper, 50\% of which is spent within the kernel.
Figure~\ref{fig:spec2006time} also shows a broken-out plot for the child
process to highlight that less than 1\% of the execution time is spent
executing decrypted instruction in user-land.

Figure~\ref{fig:execwrappertime} presents a breakdown of the execution times
within the execution wrapper while managing the full \Hydra protected program.
We evaluate the time spent generating the key using Intel's AES-CM engine,
decrypting and encrypting instructions, and writing the modified instructions
back into the child process.  Everything else within the execution wrapper
including reading from the \Hydra protected program, single-stepping the
process, logging execution, and other system miscellany are included in
\textit{Other}. Our evaluation shows that key generation, decryption, and
encryption represent negligible executions times, while writing decyrpted and
encrypted instructions back into the child process accounts for the largest
observed execution time.

\medskip \noindent\textbf{Discussion:} From a hardware perspective,
\Hydra~requires that a decryption engine decrypts instructions as they are
loaded to L1 I-cache so that no plain-text instructions are resident in the
shared instuction and data cache memory. A storage element is also necessary to
keep the decryption secrets for basic blocks. During the execution of a
process, the decryption engine is active decrypting instructions as they are
loaded by the CPU. When the CPU is executing in supervisor mode, the decryption
engine can remain inactive if the kernel code is not encrypted. Otherwise, a
secondary decryption engine can be used for code that runs in priviledged mode.
During task switching, the operating system is responsible for backing up and
restoring the decryption engine's state and associated secrets. Recent research
has demonstrated the feasibility of implementing this in hardware with limited
additional hardware~\cite{papadogiannakis2013asist}.

With hardware support the decryption engine would be placed in between the
instruction cache and CPU front-end, ensuring that instructions would remain
encrypted in the instruction cache while executing. Importantly, this would
eliminate the overhead due to both writing decrypted and encrypted instructions
back into the child process and overhead bundled into the category
\textit{Other}.

\section{Related Work}\label{sec:related}
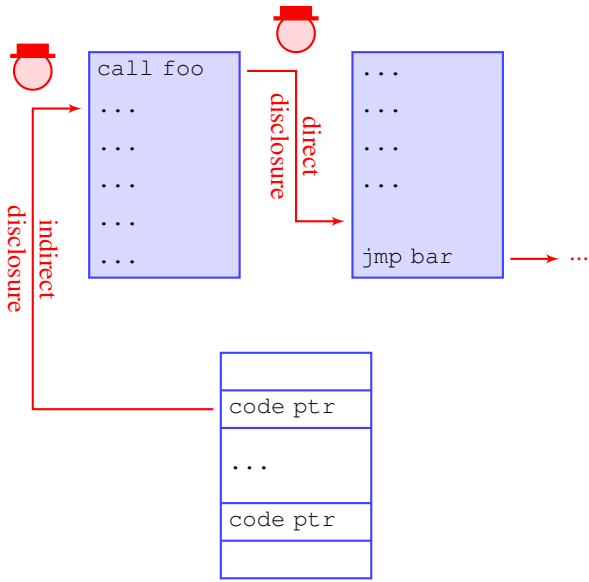
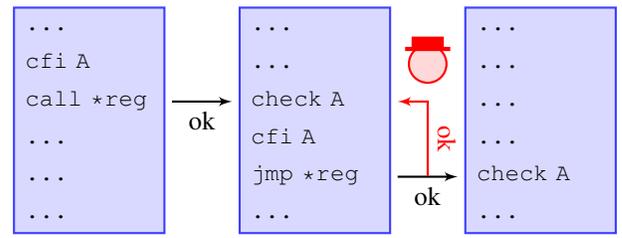
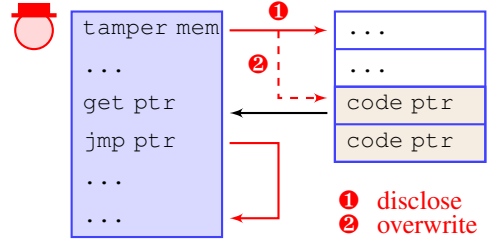
\begin{figure*}[!t]
	\centering
	\begin{subfigure}{0.40\textwidth}
	\centering
		\begin{tikzpicture}[thick, node distance=3.5cm]
	\tikzstyle{page} = [
		draw,
		rectangle,
		text width=2cm,
		minimum height=3cm,
		inner sep=0cm,
		color=blue!75,
		fill=blue!15,
		text centered,
		text=black,
	];

	\tikzstyle{data} = [
		text width=2cm,
		inner sep=0cm,
		font=\ttfamily\small,
	];




	\node[style=page] (page0) {};
	\node[style=page, right of=page0] (page1) {};
	\node[style=page, fill=white] (page2) at(1.75, -4) {};

	\node[style=data, anchor=base] at(0.1, -0.05 + 0.25 + 1) {call foo};
	\node[style=data, anchor=base] at(0.1, -0.05 + 0.25 + 0.5) {...};
	\node[style=data, anchor=base] at(0.1, -0.05 + 0.25 + 0.) {...};
	\node[style=data, anchor=base] at(0.1, -0.05 + 0.25 + -0.5) {...};
	\node[style=data, anchor=base] at(0.1, -0.05 + 0.25 + -1) {...};
	\node[style=data, anchor=base] at(0.1, -0.05 + 0.25 + -1.5) {...};

	\node[style=data, anchor=base] at(3.6, -0.05 + 0.25 + 1) {...};
	\node[style=data, anchor=base] at(3.6, -0.05 + 0.25 + 0.5) {...};
	\node[style=data, anchor=base] at(3.6, -0.05 + 0.25 + 0.) {...};
	\node[style=data, anchor=base] at(3.6, -0.05 + 0.25 + -0.5) {...};
	\node[style=data, anchor=base] at(3.6, -0.05 + 0.25 + -1.5) {jmp bar};

	\draw[color=blue!75] (-1 + 1.75, -3.5) -- (1 + 1.75, -3.5);
	\draw[color=blue!75] (-1 + 1.75, -3) --   (1 + 1.75, -3);
	\draw[color=blue!75] (-1 + 1.75, -4.5) -- (1 + 1.75, -4.5);
	\draw[color=blue!75] (-1 + 1.75, -5) --   (1 + 1.75, -5);
	\node[style=data, anchor=base] at(0.1 + 1.75, -0.05 + 0.25 + -3.5) {code ptr};
	\node[style=data, anchor=base] at(0.1 + 1.75, -0.05 + 0.25 + -4.25) {...};
	\node[style=data, anchor=base] at(0.1 + 1.75, -0.05 + 0.25 + -5) {code ptr};

	\draw[red,-latex'] (1.1, 1 + 0.25) -- (1.75, 1 + 0.25) -- 
			node[rotate=-90, text width=1.5cm, text centered]
			{direct disclosure} (1.75, 0.25 + -1) -- (2.4, 0.25 + -1);
	
	\draw[red,-latex'] (-1.1 + 1.75, 0.25 + -3.5) -- (-1.75, 0.25 + -3.5) --
			node[rotate=-90, text width=1.5cm, text centered]
			{indirect disclosure}
			(-1.75, 0.25 + 0.5) -- (-1.1, 0.25 + 0.5);

	\draw[red,-latex'] (4.6, 0.25 + -1.5) -- (5.25, 0.25 + -1.5)
			node[anchor=west] {...};

	\draw[red, fill=red!15] (-0.25+-1.5, 1.25) circle(0.25);

	\draw[red, fill=red, ultra thick] (-0.25+-1.2, 0.5+0.95) -- (-0.25+-1.8, 0.5+0.95);
	\draw[red, fill=red] (-0.25+-1.3, 0.95+0.5) rectangle (-0.25+-1.7, 0.5+ 1.1);

	\draw[red, fill=red!15] (3.25+-1.5, 1.25 + 0.5) circle(0.25);

	\draw[red, fill=red, ultra thick] (3.25+-1.2, 0.5+0.95+0.5) -- (3.25+-1.8, 0.5+0.5+0.95);
	\draw[red, fill=red] (3.25+-1.3, 0.95+0.5+0.5) rectangle (3.25+-1.7, 0.5+ 0.5+1.1);
\end{tikzpicture}
		\caption{Using both direct and indirect disclosures, software
		diversity approaches can be bypassed.}
        \label{fig:div-problem}
	\end{subfigure}
    \hfill
    \begin{minipage}{0.49\textwidth}
	\begin{subfigure}{\textwidth}
	\centering
		\begin{tikzpicture}[node distance = 3cm, thick]
	\tikzstyle{bblock} = [
		draw,
		color=blue!75,
		fill=blue!15,
		text width=2cm,
		inner sep=0cm,
		text centered,
		minimum height=3cm,
	];
	\tikzstyle{insn} = [
		text width=1.9cm,
		inner sep=0cm,
		text=black,
		font=\ttfamily\small,
	];

	\node[style=bblock] (bb0) {};
	\node[style=bblock, right of=bb0] (bb1) {};
	\node[style=bblock, right of=bb1] (bb2) {};

	\node[style=insn, anchor=base] at(0.1, -0.05 + 1.25) {...};
	\node[style=insn, anchor=base] at(0.1, -0.05 + 0.75) {cfi A};
	\node[style=insn, anchor=base] at(0.1, -0.05 + 0.25) {call *reg};
	\node[style=insn, anchor=base] at(0.1, -0.05 + -0.25) {...};
	\node[style=insn, anchor=base] at(0.1, -0.05 + -0.75) {...};
	\node[style=insn, anchor=base] at(0.1, -0.05 + -1.25) {...};

	\node[style=insn, anchor=base] at(3.1, -0.05 + 1.25) {...};
	\node[style=insn, anchor=base] at(3.1, -0.05 + 0.75) {...};
	\node[style=insn, anchor=base] at(3.1, -0.05 + 0.25) {check A};
	\node[style=insn, anchor=base] at(3.1, -0.05 + -0.25) {cfi A};
	\node[style=insn, anchor=base] at(3.1, -0.05 + -0.75) {jmp *reg};
	\node[style=insn, anchor=base] at(3.1, -0.05 + -1.25) {...};

	\node[style=insn, anchor=base] at(6.1, -0.05 + 1.25) {...};
	\node[style=insn, anchor=base] at(6.1, -0.05 + 0.75) {...};
	\node[style=insn, anchor=base] at(6.1, -0.05 + 0.25) {...};
	\node[style=insn, anchor=base] at(6.1, -0.05 + -0.25) {...};
	\node[style=insn, anchor=base] at(6.1, -0.05 + -0.75) {check A};
	\node[style=insn, anchor=base] at(6.1, -0.05 + -1.25) {...};

	\draw[-latex'] (1.1, 0.25) -- node[anchor=north] {ok} (1.9, 0.25);
	\draw[-latex'] (4.1, -0.75) -- node[anchor=north] {ok} (4.9, -0.75);
	\draw[color=red,-latex'] (4.5, -0.74) -- node[anchor=south, rotate=-90] {ok}
		(4.5, 0.25) -- (4.1, 0.25);

	\draw[red, fill=red!15] (4.5, 0.75) circle(0.25);

	\draw[red, fill=red, ultra thick] (4.2, 0.95) -- (4.8, 0.95);
	\draw[red, fill=red] (4.3, 0.95) rectangle (4.7, 1.1);

\end{tikzpicture}
		\caption{Relaxations in the policy make CFI approaches bypassable.}
        \label{fig:cfi-problem}
	\end{subfigure}
	\par\par
	\vspace*{0.2in}
	\begin{subfigure}{\textwidth}
	\centering
		\begin{tikzpicture}[node distance = 3cm, thick]
	\tikzstyle{bblock} = [
		draw,
		color=blue!75,
		fill=blue!15,
		text width=2cm,
		inner sep=0cm,
		text centered,
		minimum height=3cm,
	];
	\tikzstyle{data} = [
		draw,
		color=blue!75,
		text width=2cm,
		inner sep=0cm,
		text centered,
		minimum height=2cm,
	];
	\tikzstyle{insn} = [
		text width=1.9cm,
		inner sep=0cm,
		text=black,
		font=\ttfamily\small,
	];
	\tikzstyle{actiondesc} = [
		text width=1.9cm,
		inner sep=0cm,
		text=red,
	];

	\node[style=bblock] (bb0) {};
	\draw[
		color=blue!75,
		fill=brown!15,
	] (2 + 0.5, 0.5) rectangle (4 + 0.5, -0.5);
	\node[style=data] at(3 + 0.5, 0.5) {};

	\node[style=actiondesc] at(3 + 0.5, -1) {\bcOne~~disclose};
	\node[style=actiondesc] at(3 + 0.5, -1.33) {\bcTwo~~overwrite};

	\node[style=insn, anchor=base] at(0.1, -0.05 + 1.25) {tamper mem};
	\node[style=insn, anchor=base] at(0.1, -0.05 + 0.75) {...};
	\node[style=insn, anchor=base] at(0.1, -0.05 + 0.25) {get ptr};
	\node[style=insn, anchor=base] at(0.1, -0.05 + -0.25) {jmp ptr};
	\node[style=insn, anchor=base] at(0.1, -0.05 + -0.75) {...};
	\node[style=insn, anchor=base] at(0.1, -0.05 + -1.25) {...};

	\node[style=insn, anchor=base] at(3.1 + 0.5, -0.05 + 1.25) {...};
	\node[style=insn, anchor=base] at(3.1 + 0.5, -0.05 + 0.75) {...};
	\node[style=insn, anchor=base] at(3.1 + 0.5, -0.05 + 0.25) {code ptr};
	\node[style=insn, anchor=base] at(3.1 + 0.5, -0.05 + -0.25) {code ptr};

	\draw[color=blue!75] (2 + 0.5, 1) -- (4 + 0.5, 1);
	\draw[color=blue!75] (2 + 0.5, 0.5) -- (4 + 0.5, 0.5);
	\draw[color=blue!75] (2 + 0.5, 0) -- (4 + 0.5, 0);

	\draw[red, -latex'] (1.1, 1.25) -- node[anchor=south] {\bcOne} (2.4, 1.25);
	\draw[red, dashed, -latex'] (1.75, 1.25) -- node[anchor=east] {\bcTwo}
			(1.75, 0.25 + 0.1) -- (2.4, 0.25 + 0.1);
	\draw[latex'-] (1.1, 0.25 - 0.1) -- (2.4, 0.25-0.1);

	\draw[color=red,-latex'] (1.1, -0.25) -- (1.75, -0.25)
			-- (1.75, -1.25) -- (1.1, -1.25);

	\draw[red, fill=red!15] (-1.5, 1.25) circle(0.25);

	\draw[red, fill=red, ultra thick] (-1.2, 0.5+0.95) -- (-1.8, 0.5+0.95);
	\draw[red, fill=red] (-1.3, 0.95+0.5) rectangle (-1.7, 0.5+ 1.1);
\end{tikzpicture}
		\caption{Information hiding relies on the separation of code
		pointers from regular data. Disclosing this area of memory makes
		the approach bypassable.}
        \label{fig:ih-problem}
	\end{subfigure}
	\end{minipage}
	\caption[Bypassing mechanisms]
		{\footnotesize Different defense mechanisms against code reuse attacks and their
		bypasses. In this figure, pages shaded as
		\tikz{\draw[thick,color=blue!75,fill=blue!15] (0,0) rectangle (1.5ex,
		1.5ex);} are marked as readable and executable. Pages shaded as
		\tikz{\draw[thick, color=blue!75] (0,0) rectangle (1.5ex, 1.5ex);} are
		flagged as readable and writable. Pages shaded as \tikz{\draw[thick,
		color=blue!75,fill=brown!15] (0,0) rectangle (1.5ex, 1.5ex);} are hidden
		from the attacker, but readable and writable.}
	\label{fig:problem}
	\vspace{-0.2in}
\end{figure*}

Previous work on prevention of code-reuse attacks can be loosely categorized
into three categories: diversification, control-flow integrity, and code
pointer integrity and information hiding. While each of these mechanisms raise
the bar for attackers, they each have their limitations.  We first discuss the
merits and weaknesses of each approach using Figure~\ref{fig:problem} as a
base.

\medskip \noindent\textbf{Software diversification.} Software diversification
offers probabilistic protection by randomizing the program segments of an
application.  Code-reuse attacks are hindered by this defense because the
attacker is forced to guess code locations.  Diversification of program
segments ranges from randomizing the base offset of code and
data~\cite{team2003pax}, to shuffling the location of
functions~\cite{crane2015readactor}, dynamic sections~\cite{crane2015trap},
basic blocks~\cite{wartell2012binary, davi2013gadge}, or
instructions~\cite{hiser2012ilr, pappas2012smashing}.  These approaches can be
applied statically at compile time or dynamically at load-time or during
program execution~\cite{davi2015isomeron, bigelow2015timely}. 

Software diversification defenses, however, are only as strong as the
diversification coverage. As shown in Figure~\ref{fig:div-problem}, completely
hiding the code layout of a program has proven difficult because there are many
direct and indirect references to code locations in memory. In
JIT-ROP~\cite{snow2013just}, a single direct reference to a code location
allowed disclosure and disassembly of a significant number of code pages with
fine-grained diversification applied to them. This technique was later shown to
be applicable using indirect code pointers~\cite{davi2015isomeron}. 

\medskip \noindent\textbf{Control-flow Integrity (CFI).} CFI is a general
defense against code-reuse attacks~\cite{abadi2005control} that mitigates
control-flow hijacking by constraining execution to a legitimate control-flow
path. It does so by checking that each control-flow transfer targets a valid
code location as determined by the program's control-flow graph (CFG). The
accuracy of the statically, or dynamically, computed CFG determines the
precision of the CFI policy. This largely determines the granularity with which
control-flow checks can be made and the policy's resilience to attack.

CFI solutions in software, however, typically relax the CFG coverage in favor
of performance resulting in so-called coarse-grained
CFI~\cite{zhang2013control, zhang2013practical}. Previous work has demonstrated
that relaxation of CFG coverage leaves an application susceptible to
attack~\cite{checkoway2010return, goktas2014out, davi2014stitching}.
Figure~\ref{fig:cfi-problem} illustrates this point. The runtime address for
the register-indirect call cannot be computed statically, therefore the policy
has instrumented all targets with the same checking routine. An adversary can
redirect control at to any checking routine with the same semantics; per
Figure~\ref{fig:cfi-problem} to any target that checks label~\texttt{A}. Recent
work also calls into question the protection offered by ideally fine-grained
CFI defenses, e.g., a CFI policy supported by a completely precise CFG, using a
technique called control-flow bending~\cite{carlini2015control}. 

\medskip \noindent\textbf{Code-Pointer Integrity (CPI) and Information Hiding.}
CPI and information hiding defenses prevent writing or reading to sensitive
memory regions. This ranges from hiding code pointers in safe memory with
strict access control~\cite{kuznetsov2014code}, to hiding them behind access
permissions~\cite{backes2014oxymoron, crane2015readactor}. This defense affords
attackers limited knowledge of the vulnerable application's address space
layout and limits arbitrary manipulation of code pointers. A recent
approach~\cite{backes2014you}, called execute-no-read (XnR) delimits access
permission on code pages, marking them as executable-only. A hardware supported
version of this defense successfully mitigates a wide-range of CRAs that rely
on direct or indirect memory disclosure. Its software implementation, however,
is still susceptible to attack~\cite{werner2016no}.

Much like software diversity and CFI, information hiding has been shown to be
vulnerable to attack.  Figure~\ref{fig:ih-problem} illustrates a general
approach to bypassing information hiding. An adversary first gains control of
an arbitrary read or write vulnerability, which then is used to find and
corrupt hidden regions.  Later references to the corrupted portions of hidden
memory result in unchecked control-flow hijacking because the system assumes
derefencing code pointers from the hidden region is safe.
JIT-ROP~\cite{snow2013just} was shown to successfully bypass
Oxymoron~\cite{backes2014oxymoron}, a defense that hides direct code pointers
using Intel's segment selector registers, by leaking indirect code pointers on
virtual tables. An attack was also demonstrated against CPI by leaking its safe
memory region using timing side-channels~\cite{evans2015missing}.

\medskip \noindent\textbf{Related Approaches.} Besides instruction-set
randomization (see Section~\ref{sec:isr-related}), another work that bears some
resemblance to ours is Isomeron~\cite{davi2015isomeron}. Isomeron tolerates
complete code memory disclosure using clones (isomers) of functions within a
program. A cloned function is first diversified and then, on each control-flow
instruction, a ``coin-flip'' decides which control-flow target will be
executed, the original or clone. As the attacker is unaware of which code will
execute, crafting a control-flow hijacking attack becomes unreliable. In that
work, it is mentioned that combining ISR with diversification incurs an
unacceptable overhead.  However, we have shown that the runtime overhead of our
approach is similar on average to Isomeron, which incurs a 19\% performance
increase on average. As we have demonstrated, the main contribution to the
overhead of our system is due to the execution wrapper, which is a prototype
implementation. With full hardware decryption support, we would incur overhead
due mainly to our code diversification protection. We also protect against a
larger class of CRAs~\cite{snow2013just,carlini2015control,gawlik2016enabling}
than Isomeron.

A recent diversification approach from Crane et al.~\cite{crane2015readactor},
called Readactor, protects applications from direct and indirect memory
disclosure by combining fine-grained diversity with code-pointer hiding and
execute-no-read (XnR) protection. They achieve a practical runtime overhead of
only 6.4\% and demonstrate a working implementation by protecting the entire
Google Chromium browser and its V8 JIT compiler. Our work shares many
similarities, in particular we incorporate fine-grained code diversification
and support code hiding techniques using encryption of code pages. While we
incur a higher overhead, a key feature of our work, as opposed to Readactor, is
its resilience to CROP~\cite{gawlik2016enabling} style attacks that infer code
locations using crash-resistant memory oracles. As discussed in
Section~\ref{sec:security}, our approach withstands such an attack as it is
infeasible to disclose the plaintext and construct a reliable payload under our
code diversification and key permutation scheme even if an adversary has
knowledge of code locations.

\section{Conclusion}
In this paper, we introduce \emph{\Hydra}, the first hybrid defense
incorporating software diversity, control-flow integrity, and information
hiding. We demonstrate how an encryption chain over basic blocks can be used to
achieve \emph{execution integrity}. When combined with software diversification
techniques and information hiding, \Hydra's per-basic-block encryption yields a
mechanism that is capable of protecting against a diverse set of code-reuse
attacks under a powerful attacker model. We also show the applicability of our
mechanism to a multitasking system while demonstrating reasonable performance
overhead when benchmarked with the SPEC CPU2006 suite.\par

\bibliographystyle{IEEEtran}
\bibliography{deanbib}
\end{document}